# Patterns and Thresholds of Magnetoelectric Switching in Spin Logic Devices

Dmitri E. Nikonov[a)], Sasikanth Manipatruni, and Ian A. Young

*Components Research, Intel Corp., Hillsboro, Oregon, 97124, USA*

In the quest to develop spintronic logic, it was discovered that magnetoelectric switching results in lower energy and shorter switching time than other mechanisms. Magnetoelectric (ME) field due to exchange bias at the interface with a multi-ferroic (such as $BiFeO_3$) is well suited for 180 degree switching of magnetization. The ME field is determined by the direction of canted magnetization in $BiFeO_3$ which can point at an angle to the plane, to which voltage is applied. Dependence of switching time and the threshold of ME field on its angles was determined by micromagnetic simulations. Switching occurs by formation of a domain wall on the side of the nanomagnet on top of BFO and its propagation to the rest of the magnet. For in-plane magnetization, switching occurs over a wide range of angles and at all magnitudes of ME field above threshold. For out-of-plane magnetization failure occurs (with an exception of a narrow range of angles and magnitudes of ME field) due to the domain wall reflecting from the opposite end of the nanomagnet.

**I. INTRODUCTION**

Over the past five decades, complimentary metal-oxide-semiconductor (CMOS) field effect transistors (FET) enabled a revolution in computing via continuous scaling of their sizes according to the Moore's law[1]. Presently significant research effort in the field is devoted to discovery of beyond CMOS computing hardware[2], including spintronics, i.e. nanomagnet based devices. Benchmarking of beyond CMOS device and circuit options[3] led to a conclusion that magnetoelectric effects is the most promising method of switching spintronic devices.

Various mechanism of magnetoelectric switching exist: surface anisotropy (aka voltage controlled magnetic anisotropy, VCMA)[4], magnetostrictive switching[5,6], and exchange bias[7,8]. All of them have been experimentally demonstrated[9,10,11,12]. Exchange bias stands out among the three since it has a definite direction determined by voltage. Thus it is able to reverse magnetization and give it a direction determined by voltage regardless of its prior state. The other two mechanisms create a preferred axis for magnetization, but not a definite direction along this axis. MS and VCMA mechanisms are still able to reverse magnetization[13], but require an external magnetic field bias. However in this case the final magnetization is opposite to the initial one, rather than being determined by the voltage input.

---

[a]Author to whom correspondence should be addressed. Electronic mail: Dmitri.e.nikonov@intel.com.

In view of the advantage of exchange bias, logic devices based on it were proposed, such as magnetoelectric-spin-orbital (MESO) logic[14]. It envisioned switching by exchange bias from multiferroic BFO ($BiFeO_3$). Experimentally, the magneto-electricity from multiferroic $BiFeO_3$ is realized in Rhombohedral $BiFeO_3$ (100) textured substrates grown on perovskite substrates ($DyScO_3$ or $SrTiO_3$). Rhombohedral-$BiFeO_3$ (100) is a G-type AFM with alternate FM planes formed by $Fe^{3+}$ in the [111] crystal plane. This creates a net magneto-electric effect oriented at an angle to the top surface of BFO. Important geometry considerations arise in MESO devices due to: a) effective ME field due to exchange bias being in general at a certain angle to the interface with BFO; b) this interface extending only over a part of a nanomagnet. It is desirable to anticipate consequences of such geometry factors for device performance by simulating them.

Most of simulations of magnetization dynamics in magnetoelectric switching has been performed macrospin (i.e., uniform magnetization) approximation, e.g.[15]. Only a few publications contain micromagnetic (i.e. non-uniform magnetization) results, e.g.[16]. We expect magnetoelectric switching to be strongly influenced by magnetization patterns. For benchmarking of ME devices, it is important to predict the threshold values of ME field for the onset of switching. Especially for nanoscale magnets, these values may be very different than those observed in experiments with micron-size samples. In this paper we address the above issues by micromagnetic simulations, make conclusions about the patterns of magnetoelectric switching, and determine switching thresholds and their angle dependence for a few characteristic device geometries.

## II. GEOMETRY OF MAGNETOELECTRIC SWITCHING

Exchange bias at the interface of multiferroic magnetoelectric BFO has both the switching scenario and geometry which are non-trivial, see Ref.7. In Rhombohedral BFO, the spontaneous ferroelectric polarization (P), pointing at a corner of a cubic crystal lattice cell (also the rhombohedral axis), is switched by applying electric field (E). Polarization is tied to the antiferromagnetic order (L), and to the net canted magnetization (Mc=$M_1$-$M_2$, where $M_1$, $M_2$ are the magnetizations of the sub lattices of BFO) which is perpendicular to P and L. For example if E is applied perpendicular to the plane of the device and has (0,0,-1) crystal direction and P is (1,-1,1), then L can be (1,1,0) and Mc be (1,-1,-2), which is ~35.5º from the z-axis (perpendicular to the plane). Also switching of P in response to E happens in two or three consecutive stages where the polarization P consecutively rotates along the [±1,±1,±1] directions till a complete reversal is achieved. Note that the polarization does not go through a zero polarization state. The exchange bias (equivalently ME field, $H_{ME}$) is related in an as yet unknown way to in-plane and out-of-plane projections of Mc. Additionally, by device design, the long axis of the nanomagnet can be positioned at a certain angle to the crystal axes. In summary, the magneto-electric switching in multiferroic 100 oriented Rhombohedral-BiFeO3 has the following unique properties a) the polarization switching in Rhombohedral-BFO under goes a non-Vanderbilt switching (i.e., polarization never goes to zero) where the Polarization vector rotates from one body diagonal direction to another till a 180º reversal is obtained. b) The magnetization and L vector always point along the plane perpendicular to the P and also under go 180º reversal. c) The net magnetization Mc=$M_1$-$M_2$, where $M_1$, $M_2$ are the magnetizations of the sub lattices of BFO) points out of plane as dictated by the dynamics of P.

We can explore the whole range of angles of the ME field, specified in terms of the angle to the long axis and to the device plane (Figure 1). We simulate both the cases of ME field applied to the whole nanomagnet ('whole field') as well as only over the 'write area' ('section field'), on the left of the nanomagnet as in Figure 1. In the last case, magnetization can be read off in the 'read area' on the right of the nanomagnet via inverse spin orbit coupling effect, as in Ref.14.

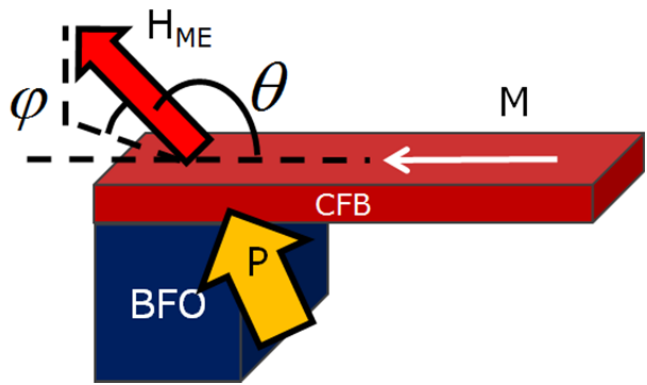

FIG. 1. Scheme of a magnetoelectric device switched by ME field. Nanomagnet consists of CoFeB alloy and has magnetization M. Multiferroic BiFeO3 has voltage-switched polarization P. Angles of ME field with the long x-axis of the nanomagnet is θ, and with the device plane is φ. For out-of-plane magnetization, the angles are related to the z-axis.

### III. MICROMAGNETIC SIMULATIONS AND DOMAIN WALL FORMATION

We use a standard micromagnetic solver OOMMF[17] to model magnetization dynamics in a nanomagnet. Material parameters approximating CoFeB alloy are taken: magnetization $M_s = 1.3 MA/m$, exchange stiffness $A_{ex} = 12 pJ/m$, Gilbert damping $\alpha = 0.1$. We chose the sizes of the magnets to ensure the energy barrier of at least $60 k_B T$. Specifically for in-plane magnetization the size is 60x20x2.7nm. For the whole ME field, write and read areas are 20nm long each. For out-of-plane magnetization, it is ensured by surface anisotropy $K_i = 1.2 mJ/m^2$. The nanomagnet size is 60x20x1nm for the whole ME field. The size is 100x20x1nm for the section ME field, while the write area is 60nm and the read area is 20nm long. For the whole ME field, switching happens with close to uniform magnetization distribution and this pattern remains similar for a range of magnitude of $H_{ME}$. The situation is very different for the section ME field cases. In in-plane nanomagnets, Figure 2, switching starts in the write area; a domain wall forms, it propagates to the opposite edge, and then disappears off the edge.

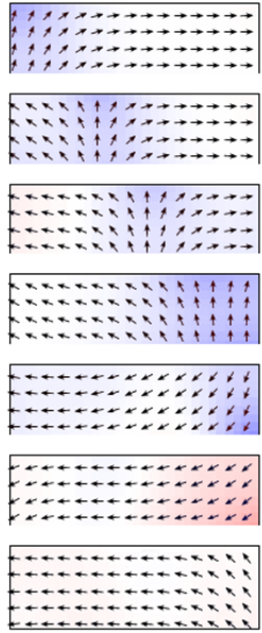

FIG. 2. Snapshots of magnetization in the in-plane nanomagnet at 30ps intervals from application of $H_{ME}$=2000 Oe with angles $\theta$=160$^o$ and $\varphi$=40$^o$. Red designates positive M projection on the z-axis, blue designates negative.

Similar propagation and disappearance happens in out-of-plane nanomagnets for certain combinations of ME field magnitude and angles. However in the majority of cases, dynamics is different. The domain wall propagates toward the opposite edge, reflects off this edge, sometimes moves over a certain distance and then stops, see Figure 3. Similar dynamics was observed in domain walls driven by spin transfer torque[18]. We consider it a failure of switching and a strong reason to avoid out-of-plane nanomagnet designs.

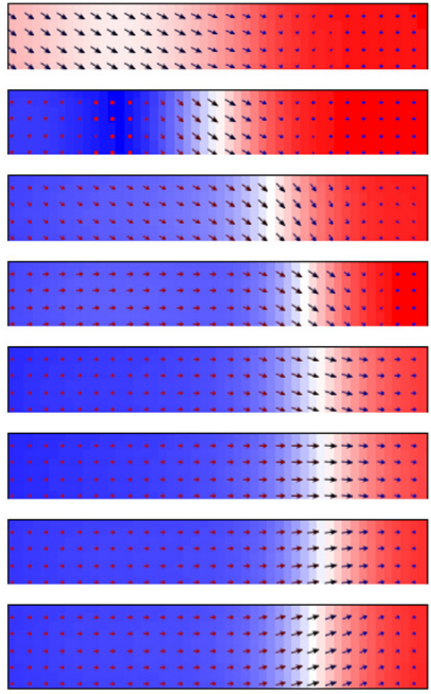

FIG. 3. Snapshots of magnetization in the out-of-plane nanomagnet at 50ps intervals from application of $H_{ME}$=2000 Oe with angles $\theta$=120° and $\varphi$=0°.

**IV. THRESHOLDS OF SWITCHING**

Simulations as described above were run ran over a wide range of angles and magnitudes of ME field. We aim at calculating the switching speed, which is defined as inverse of the switching time – from the onset of the ME field and relative magnetization of +1 in the nanomagnet to the last time relative average magnetization reaches -0.5 in the read area.

Switching is well behaved in the whole ME field case for either in-plane or out-of-plane nanomagnets: problems with domain wall reflection are not present. The dependence of switching speed on ME field magnitude shows a sharp threshold and then linear increase of speed, looking similar to that in Figure 4. The threshold is a relatively slowly changing function on angle.

Minimal value of threshold for in-plane magnetization is 700 Oe at angles $\theta$=130° and $\varphi$=0°. For out-of-plane magnetization it is 700 Oe at angles $\theta$=120° and $\varphi$=20°.

These trends remain qualitatively the same in section ME field on an in-plane nanomagnet, shown in Figure 4. Quantitatively, the switching speed is ~2.5 slower, and the threshold is higher, 1600 Oe at angles $\theta$=160° and $\varphi$=40°. The threshold most strongly depends on $\theta$ and weakly on $\varphi$.

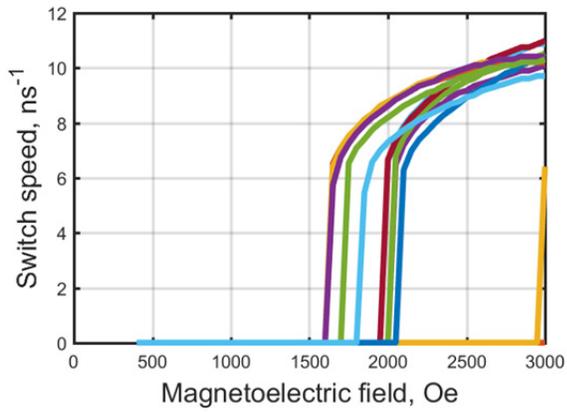 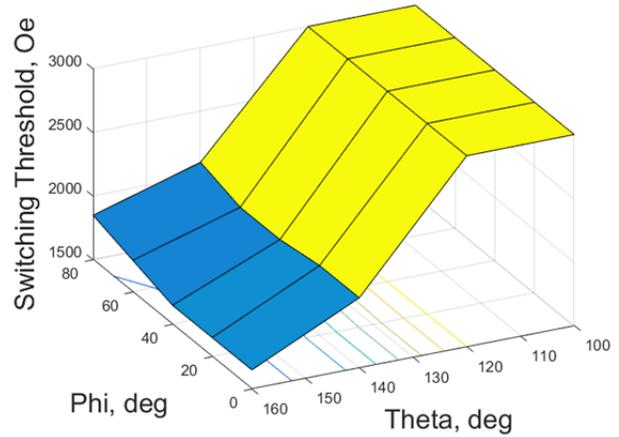

FIG. 4. Magneto-electric switching for in-plane magnetization. a) Switching speed vs. ME field magnitude, curves correspond to the range of angles in (b). b) Threshold ME field vs. angles.

The trends are quite different for the section ME field and out-of-plane nanomagnets – due to domain wall reflections, as noted above, see Figure 5. Dependence on the ME field magnitude shows a threshold but then switching breaks down at higher ME fields. Dependence on ME angles is also much sharper, especially on θ (note a much more expanded scale). Switching fails outside the range of angles shown in the plot. Since the switching intervals vs. ME field parameters are small, this geometry option is not robust enough for practical devices.

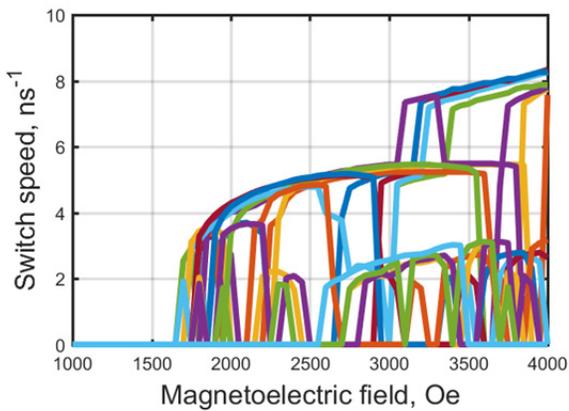 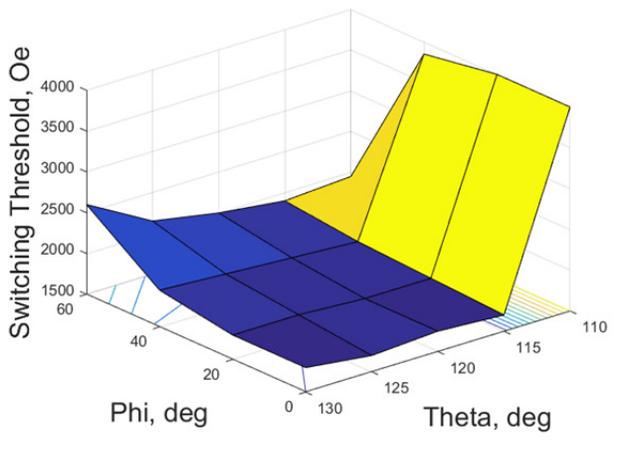

FIG. 5. Magnetoelectric switching for out-of-plane magnetization. a) Switching speed vs. ME field magnitude, curves correspond to the range of angles in (b). b) Threshold ME field vs. angles.

## V. CONCLUSION

We study the micro magnetic of magneto-electric switching from multi-ferroic Rhombohedral BiFeO$_3$ (100) textured substrates, comprehending the out of plane components of the ME field. We performed micro-magnetic simulations of exchange bias switching with relevant geometrical aspects. We summarize that: a) Well behaved switching of magnetization occurs for ME field applied over the whole nanomagnet for both in plane and out of plane magnets b) For ME field applied over a part of the nanomagnet (a natural requirement in several devices), switching occurs via formation of domain walls. c) In plane magnets provide better switching characteristics for ME field applied over a part of the nanomagnet than an out of plane magnet does. Problems arise when domain walls reflect in nanomagnets with out-of-plane magnetization. Acceptable switching behavior occurs for the case of in-plane magnetization with typical threshold ME field of 1600 Oe.

## ACKNOWLEDGMENTS

The authors would like to thank R. Ramesh and S. Salahuddin for inspiring discussions.

## REFERENCES

[1] G. E. Moore, Electronics, 38, no. 8, 114-117 (1965).
[2] D. E. Nikonov and I. A. Young, Proc. IEEE, 101, no. 12, 2498-2533 (2013).
[3] D. E. Nikonov and I. A. Young, IEEE J. Explor. Comput. Devices and Circuits 1, 3-11 (2015).
[4] F. Matsukura, Y. Tokura and H. Ohno, Nature Nanotech., 10, 209 (2015).
[5] W. Eerenstein, N. D. Mathur, and J. F. Scott, Nature, 442, 759 (2006).
[6] M. Fiebig, J. Phys. D: Appl. Phys., 38, R123–R152 (2005).
[7] J. T. Heron et al., Nature, 516, 370 (2014).
[8] C. A. F. Vaz , J. Hoffman , C. H. Ahn , and R. Ramesh, Advanced Materials, 22, 2900–2918 (2010).
[9] T. Maruyama et al., Nat. Nanotechnol. 4, 158 (2009).
[10] X. He, Y. Wang, N. Wu, A. N. Caruso, E. Vescovo, K. D. Belashchenko, P. A. Dowben, and C. Binek, Nature Mater. 9, 579 (2010).
[11] Y. Chen, T. Fitchorov, C. Vittoria, and V. G. Harris, Appl. Phys. Lett. 97, 052502 (2010).
[12] J. T. Heron et al., Phys. Rev. Lett. 107, 217202 (2011).
[13] P. Khalili Amiri et al., IEEE Transactions on Magnetics, 51, 1-7 (2015).
[14] S. Manipatruni, D. E. Nikonov, and I. A. Young, "Spin-Orbit Logic with Magnetoelectric Nodes: A Scalable Charge Mediated Nonvolatile Spintronic Logic", http://arxiv.org/abs/1512.05428.
[15] A. Sukhov, C. Jia, P. P. Horley, and J. Berakdar, J. Phys.: Condens. Matter 22 (2010).
[16] Z. Wang and M. J. Grimson, J. Appl. Phys., 119, 124105 (2016).
[17] M. J. Donahue and D. G. Porter, National Institute of Standards and Technology Report No. NISTIR 6376, September 1999.
[18] D. E. Nikonov, S. Manipatruni, and I. A. Young, J. Appl. Phys. 115, 213902 (2014).